\documentstyle[12pt,fleqn,epsfig]{article}
\begin{document}

\title{Open charm enhancement by secondary interactions in relativistic
nucleus-nucleus collisions?\thanks{supported by DFG/RFFI and the
Landau-Heisenberg program} }
\author{W. Cassing$^1$, L. A Kondratyuk$^2$, G. I. Lykasov$^3$, M. V.
Rzjanin$^3$ \\ {\normalsize $^1$Institut f\"{u}r Theoretische
Physik,}\\ {\normalsize Universit\"{a}t Giessen,} {\normalsize
35392 Giessen, Germany} \\ {\normalsize $^2$Institute of
Theoretical and Experimental Physics,} \\ {\normalsize B.
Cheremushkinskaya 25, 1172259 Moscow, Russia} \\ {\normalsize
$^3$Joint Institute for Nuclear Research,} \\ {\normalsize 141980
Dubna, Moscow Region, Russia }}

\date{}

\maketitle

\begin{abstract}
We calculate open charm production in $Pb+Pb$ reactions at SPS
energies within the HSD transport approach - which is based on
string, quark,  diquark ($q, \bar{q}, qq, \bar{q}\bar{q}$) and
hadronic degrees of freedom - including the production of open
charm pairs from secondary  'meson'-'baryon' (or quark-diquark and
antiquark-diquark) collisions. It is argued that at collision
energies close to the $c\bar{c}$ pair threshold the dominant
production mechanism is related to the two body (or quasi two
body) reactions $\pi N \rightarrow \bar{D} (\bar{D^*}) \Lambda_c,
(\Sigma_c)$. Estimates within the framework of the Quark-Gluon
String model suggest cross sections of a few $\mu b$ for $\pi N
\rightarrow \bar{D} \Lambda_c$  in  the region of 1 GeV above
threshold. The dynamical transport  calculations for $Pb + Pb$ at
160 A$\cdot$GeV indicate that the open charm  enhancement reported
by the NA50 Collaboration might be due to such secondary reaction
mechanisms.
\end{abstract}

\vspace{1cm} \noindent PACS: 25.75.-q; 13.60.L2; 14.40.Lb; 14.65.Dw


\newpage

\section{Introduction}

In the last decade the interest in hadronic states with charm
flavors  ($c, \bar{c}$) has been rising continuously in line with
the development of  new experimental facilities
\cite{QM99}. This relates to the charm production cross
section in $pN$ and $\pi N$ reactions as well as to  their
interactions with baryons and mesons which determine their
properties (spectral functions) in the hadronic medium. The charm
quark degrees of freedom play an important role especially in the
context of a phase transition to the quark-gluon plasma (QGP)
\cite{Heinz} where charmonium ($c\bar{c}$) states should no longer
be formed due to color screening \cite {Satz,Satznew}. However,
the suppression of $J/\Psi$ and $\Psi^\prime$ mesons in the high
density phase of nucleus-nucleus collisions \cite {NA50b,NA50a}
might also be attributed to inelastic comover scattering  (cf.
\cite{Cass99,Vogt99,Gersch,Cass00,Capella} and Refs. therein)
provided that the corresponding $J/\Psi$-hadron cross sections are
in the order of a few mb \cite{Haglin,Haglin2,Konew,Ko,Sascha2}.
Present theoretical estimates here differ by more than an order of
magnitude \cite{Bernd} especially  with respect to $J/\Psi$-meson
scattering such that the question of  charmonium suppression is
not yet settled. On the other hand, the enhancement of
'intermediate-mass dileptons' in $Pb + Pb$ collisions at the SPS
has  been tentatively attributed to an enhancement of 'open charm'
in  nucleus-nucleus collisions relative to $pA$ reactions at the
same invariant energy  $\sqrt{s}$ \cite{NA50d}. Thus 'charmonium
suppression' and 'open charm enhancement'  are present facets of
relativistic heavy-ion collisions, which provide a theoretical
\cite{Ko,Johanna,Peter,Rafelski,Wang,Lin00} and experimental
challenge for the future \cite{NA50new,NA49new}.

In this letter we will argue that the open charm excess reported
in Ref. \cite{NA50d} might be due to secondary meson-baryon
(quark-diquark) interactions. We will employ the HSD transport
approach  \cite{Cass99,Cass00,Ko95} for the overall reaction
dynamics of nucleus-nucleus collisions using parametrizations for
the elementary production channels for the charmed hadrons $D,
\bar{D}, D^*, \bar{D}^*, D_s, \bar{D}_s, D_s^*,\bar{D}_s^*,$
$J/\Psi, \Psi(2S), \chi_{2c}$ from $NN$ and $\pi N$ collisions. We
recall that in the HSD transport approach the initial stages of a
$\pi N$, $pp$, $ pA$ or $AA$ reaction (at high energy) are
described by the excitation of color neutral strings, where the
leading quarks and 'diquarks' in a  baryonic string (or quarks and
antiquarks in a mesonic string etc.) are allowed  to rescatter
again (in case of nuclear targets) with hadronic cross sections
divided by the number of constituent quarks and antiquarks in the
hadrons, respectively  \cite{Geiss}.

In order to explore the kinematical situation for secondary
'meson'-'baryon' interactions we show in Fig. 1 the distribution
in the invariant collision energy $\sqrt{s}$ from such secondary
interactions (solid histogram) for a central collision of $Pb +
Pb$ at 160 A$\cdot$GeV from the HSD approach. This distribution
extends to about 15 GeV while the corresponding primary
baryon-baryon collisions are centered around 17.3 GeV (dashed
histogram) and become less frequent than the 'meson'-'baryon'
collisions for $\sqrt{s} \leq$ 13 GeV. Since the threshold for
open charm production in meson-baryon collisions is $\sqrt{s_0} =
m_D + m_{\Lambda_c} \approx$ 4.15 GeV there are a lot of secondary
collisions above threshold. The crucial question now is the
magnitude of the open charm cross section in 'meson'-'baryon
interactions for 4.15 GeV $\leq \sqrt{s} \leq$ 15 GeV, i.e from
threshold up to roughly 10 GeV above threshold.

As argued in Refs. \cite{Peter,new} the creation of a $c\bar{c}$
pair at high $\sqrt{s}$ is due to a hard process and dominated by
gluon-gluon fusion (cf. Fig. 2a) for an illustration of a $pp$
reaction). On the other hand, the quark annihilation mechanism is
found to contribute significantly or even to become dominant for
$\pi N$ reactions at lower $\sqrt{s}$ \cite{Vogtnew}. This
mechanism is depicted in Fig. 2b) for the particular final state
$\bar{D} \Lambda_c$, where -- by convention -- $\bar{D}$ stands
for a $D$-meson with a $\bar{c}$-quark. However, the perturbative
$q \bar{q} \rightarrow 1 \ \rm{gluon} \rightarrow c\bar{c}$
process is included in standard PYTHIA calculations \cite{PYTHIA}
and drops very fast with decreasing $\sqrt{s}$ close to the
threshold of $D-$meson production \cite{new}. We recall that in
Ref. \cite{new} the open charm cross sections for $pN$ and $\pi N$
reactions have been calculated within PYTHIA using MRS G (next to
leading order) structure functions from the PDFLIB package
\cite{MRS} for the gluon distribution of the proton, a bare charm
quark mass $m_c= $ 1.5 GeV and $k_T$ =  1 GeV. The results from
PYTHIA (scaled to the available data \cite{new}) for $D, \bar{D},
D^*, \bar{D}^*, D_s, \bar{D}_s, D_s^*, \bar{D}_s^*,$ as a function
of $\sqrt{s}\geq $ 10 GeV have been parametrized by an expression
of the form,
\begin{equation}
\sigma_X(s) =  a_X (1 - Z)^\alpha \ Z^{-\beta},  \label{fit}
\end{equation}
with $Z =  {\sqrt{s^0_X}}/{\sqrt{s}}$ where $\sqrt{s^0_X}$ denotes
the threshold for the channel $X$ in $pN$ or $\pi N$ reactions.
The individual parameters $a_X, \alpha$ and $\beta$ from this fit
are given in Table 1 for $\pi N$ reactions. As found in Ref.
\cite{new} the contribution to open charm production from
'secondary' interactions -- employing the parametrizations
(\ref{fit}) -- are about $\sim$ 9 \% in central $Au + Au$
collisions at 160 A$\cdot$GeV. This order of magnitude for the
open charm enchancement is far below the observation of Ref.
\cite{NA50d}.

\begin{table}[t]
\caption{ The parameters $a_x, \alpha$ and $\beta$ in Eq. (\ref{fit}) for
inclusive $D$-meson production in $\pi N$
reactions} \label{tabl2}
\par
\begin{center}
\vspace{0.5cm}
\begin{tabular}{|lllll|}
\hline \multicolumn{5}{|c|}{$\pi N$} \\ \hline Meson &
$\sqrt{s_0}$ [GeV] & $a_x$ [mb] & $\alpha$ & $\beta$ \\ \hline
$D^0$ & 4.667 & 0.273 & 2.86 & 1.28 \\ $\bar D^0$ & 4.150 & 0.247
& 3.80 & 1.26 \\ $D^+$ & 4.671 & 0.255 & 3.22 & 1.28
\\ $D^-$ & 4.154 & 0.286 & 3.50 & 1.22 \\ \hline $D^{0*}$ &
4.951 & 1.076 & 3.14 & 1.22 \\ $\bar D^{0*}$ & 4.292 & 0.774 &
3.80 & 1.26 \\ $D^{+*}$ & 4.955 & 0.719 & 2.86 & 1.32 \\ $D^{-*}$
& 4.296 & 0.839 & 3.40 & 1.24 \\ \hline $D^+_s$ & 4.875 & 0.0932 &
3.62 & 1.22 \\ $D^-_s$ & 4.435 & 0.0545 & 3.70 & 1.34
\\ $D^{+*}_s$ & 5.162 & 0.284 & 3.42 & 1.24 \\ $D^{-*}_s$ &
4.578 & 0.163 & 3.64 & 1.34 \\ \hline
\end{tabular}
\end{center}
\end{table}

On the other hand, close to threshold, one would expect the
exclusive reaction $\pi N\rightarrow  \bar{D} \Lambda_c$ to
dominate. In case of $S$ -wave production this cross section then
should rise as $\sim  (1-Z)^{0.5}$ for a two-body final channel.
Such a behaviour is well known experimentally \cite{LB} for the
reactions $\pi^+ p \rightarrow \rho^+ p$, $\pi^- p \rightarrow
\omega n$, $\pi^- p \rightarrow K^0 \Lambda$ or $\pi ^- p
\rightarrow \phi n$ as displayed in Fig. 3. These experimental
cross sections can be described by the expression
\begin{equation}
\label{sfit}
\sigma _{\pi N\rightarrow XB}(\sqrt{s})= a_X(1-\frac{\sqrt{s_0}}{\sqrt{s}}%
)^{0.5}(\frac{\sqrt{s_0}}{\sqrt{s}})^{\gamma_{X}}  \label{pin}
\end{equation}
with $a_\rho = 15$ mb, $a_\omega = 12$ mb, $a_{K^0} = 2.5$ mb,
$a_\phi =$ 0.2 mb and $\gamma_\omega = \gamma_K$ = 6,
$\gamma_\rho$ = 5 and $\gamma_\phi$ = 10, respectively. In Eq.
(\ref{sfit}) $\sqrt{s_0}$ denotes the threshold for each reaction
separately. For $\sqrt{s}- \sqrt{s_0} \geq m_\pi$ three-body final
states (with an additional pion) become possible while
multi-particle production dominates at high invariant energy. To
demonstrate the relative importance of multi-particle final states
the inclusive cross sections for $\rho^+, \omega$ and $\phi$ are
shown in terms of the thick lines in Fig. 3 using the
parametrizations from the review \cite{Cass99}. The inclusive
yield for all open charm mesons with a $\bar{c}$ quark is shown in
terms of the lower thick solid line in Fig. 3 where the
parametrisations (\ref{fit}) have been employed with the
parameters from Table 1 (cf. \cite{new}). The related data (full
squares) have been taken from the review of Tavernier
\cite{Tavernier}. We recall that for $\sqrt{s} - \sqrt{s_0} \geq $
6 GeV the parametrization reflects the PYTHIA results (scaled to
the available data) and at lower energies lacks a more fundamental
justification as pointed out in Ref. \cite{new}.

Since the
binary exclusive reactions dominate the threshold behaviour
 for $\rho, \omega, K^0$ and $\phi$ production,
one can expect a similar relation to hold also for the channels
$\bar D\Lambda _c$, $\bar D^{*}\Lambda _c$, $\bar D\Sigma _c$,
$\bar D^{*}\Sigma _c, \ etc$. Here the reaction mechanism is
expected to be dominated by $D^*$-meson or $D^*$-Reggeon  exchange
as illustrated in Fig. 2c). The question that remains is the size
of a related parameter $a_D$ in (\ref{sfit}). Here we refer to the
Regge model for an estimate.

In Ref. \cite{Boreskov} the cross section for the reaction $\pi N
\to \bar D(\bar D^*)\Lambda_c$ has been estimated within the
framework of the Quark-Gluon String Model (QGSM) developed in Ref.
\cite{Kaidalov1}. The GGSM is a nonperturbative approach based on
a topological $1/N_f$ expansion in QCD, where $N_f$ denotes the
number of flavors, and on the color-tube model. This approach can
be considered as a microscopic model describing Regge
phenomenology in terms of quark degrees of freedom. It provides
the possibility to establish relations between many soft hadronic
reactions as well as masses and partial widths of resonances with
different quark contents (see e.g. the review
\cite{KaidalovSurveys}). Recently, this model also has been
successfully applied to the description of the nucleon and pion
electromagnetic form factors \cite{Tchekin} and deuteron
photodisintegration \cite{Grishina}.

The amplitude of the reaction $\pi N \to \bar D(\bar D^*)
\Lambda_c$ corresponding to the planar graph of Fig. 2c) with $u$
and $\bar c$ quark exchange in the $t$ channel -- using the
Mandelstam variables $s$ and $t$ -- can be written as (see Ref.
\cite{Boreskov})
\begin{equation}
A_{\pi^- p \to \bar D  \Lambda_c}(s,t)=  \sqrt{2} g^2_0
F(t)(s/s_0^{u  \bar c}) ^{\alpha_{u\bar c}(t) -1} (s/{\bar s}),
\end{equation}
where $g_0 \simeq 5.8$ is a universal flavor independent
coupling constant, $\alpha_{u\bar c}(t) =  \alpha_{D^*}(t) $ is
the $D^*$ Regge trajectory, $\bar s = 1$ GeV$^2$, $s_0^{u \bar
c}$=  4.9 GeV$^2$ is the flavor dependent scale factor and $F(t)$
is the form factor describing the $t$ dependence of the residue.
We have considered two forms of the $D^*$ Regge trajectory, i.e.
\\ i) a linear trajectory
(set $a$ from Ref. \cite{Boreskov})
\begin{equation}
\alpha_{D^*}(t)= -0.86 +0.5 t
\end{equation}
ii)a nonlinear trajectory, i.e. the square root trajectory from Ref. \cite{Burak}
\begin{equation}
\alpha_{D^*}(t)=  \alpha_{D^*}(0) - \gamma [\sqrt{T} - \sqrt{T-t}]
\end{equation}
with  $\alpha_{D^*}(0) = -1.02$, $\gamma= 3.65$ GeV$^{-1}$,
$\sqrt{T}= $3.91 GeV. Both choices lead to very similar total
cross sections such that we do not discuss the two choices
separately. The form factor $F(t)$ determining the $t$-dependence
of the residue has been parametrized in Ref. \cite{Boreskov} as
\begin{equation}
\label{Gamma} F(t)= \Gamma(1-\alpha_D{^*}(t))
\end{equation}
and is motivated by dual models.  The form factor (\ref{Gamma}) is
convenient for an analytical continuation of the amplitude to the
region of positive $t$, where the $\Gamma$ function in
(\ref{Gamma}) is an exponentially decreasing function of $t$.  In
particular, the form (\ref{Gamma}) was successfully used to relate
the normalization of different planar graphs at $t$= 0 with the
widths of the mesons lying on the corresponding Regge
trajectories. For example, in Ref. \cite{Volkovitsky} a width
$\Gamma(D^{*+} \to D^0 \pi^+ + D^+ \pi^0)= (3/2)\Gamma(D^{*+} \to
D^0  \pi^+)$=  20 keV was found. This value is not very different
from the result based on the QCD sum rules \cite{QCDsumrule}
giving $\simeq 48$ keV, while the experimental upper limit is 131
keV.

However, in the region of negative $t$ the parametrization
(\ref{Gamma}) exhibits a factorial growth (which is faster than
exponential) and therefore is not acceptable. For $t \leq 0$ we
will use the convential parametrization
\begin{equation}
\label{exp} F(t)= \Gamma(1-\alpha_D{^*}(0)) \ {\mathrm exp} (R^2
t)
\end{equation}
with $R^2=  0\div 0.2$ GeV$^{-2}$.

The differential cross section for the reaction $\pi^- p \to \bar
D  \Lambda_c $ then is
\begin{eqnarray}
\label{diff} &\displaystyle \frac{d\sigma_{\pi^- p \to \bar D
\Lambda_c }}{d t}& =  \frac{1}{64\,\pi s}\
\frac{1}{(p_{\pi}^{\mathrm{cm}})^2}\ \ | A_{\pi^- p \to \bar D
\Lambda_c}(s,t)|^2 ,
\end{eqnarray}
where $p_{\pi}^{\mathrm{cm}}$ denotes the pion momentum in the
cms, and the total cross section is obtained by integrating
(\ref{diff}) over the kinematical allowed regime.

The result for the channel $\pi^- p \rightarrow \bar{D} \Lambda_c$
(using $R^2 =0$) is shown in Fig. 3 by the lower solid line as a
function of $\sqrt{s}-\sqrt{s_0}$, which can be well described by
the function (\ref{sfit}) using $a_D$ = 0.027 mb and $\gamma_D =
\gamma_\omega = \gamma_K$ = 6 (lower dashed line). The ratio of
the $\omega$ cross section to the 'estimated' $\bar{D}$ cross
section close to threshold ( $\sqrt{s}-\sqrt{s_0} \approx 0.4$
GeV) is $\sim$ 500; a similar ratio holds for the inclusive cross
sections at $\sqrt{s}-\sqrt{s_0} \approx 30$ GeV where explicit
data are available for the $\pi N \rightarrow D \bar{D} + X$
reaction. Thus our estimate in the Regge model sounds reasonable,
however, has to be controlled by experiment in order to allow for
definite conclusions.

We mention that similar exclusive cross sections can also be
obtained from boson-exchange models with $D^*$ or $D_1$ exchange
(cf. Fig. 2c) employing monopole form factors at the vertices with
cutoffs $\Lambda \ (\approx 1.6 - 1.9$ GeV) while fixing the
couplings $g_{\bar{D} \pi \bar{D}^*}$ and $g_{\bar{D}^* \pi =
D_1}$ to the $\pi \bar{D}$ and $\pi \bar{D}^*$ decay widths,
respectively. Furthermore, the cross section for the reaction $NN
\rightarrow \bar D(D^*)\Lambda_c N$ has been calculated  within
the Reggezied one-pion exchange model using the pion form factor
$F_\pi=\exp(R_\pi^2 t)$ with $R_\pi^2=0.1(GeV/c)^{-2}$ as in Ref.
\cite{Boreskov}. Since the explicit cross sections for the $NN$
reactions at low $\sqrt{s}$ are of no major importance for our
present study, we refer the reader to a forthcoming publication on
these issues \cite{Lykasov}.

For the transport calculations (to be discussed below) we estimate
the cross section for charmed hadron production in $\pi N$
reactions by employing the ansatz (\ref{sfit}) for all charmed
hadrons such as $\bar{D},  \bar{D}^*, \bar{D}_s, \bar{D}_s^*$ with
associated charmed hyperons $\Lambda_c, \Sigma_c, \Lambda_c^s
...$. We explicitly adopt $a_D   = 1/3 a_{D^*} =  a_{D_s^*} =  3
a_{D_s} $ (as in Ref. \cite{new}) with $a_D  \approx 0.027$ mb.
These estimates, of cause, only have exploratory character as
stated above.

We now turn to the results of transport calculations for the
reaction $Pb +Pb$ at 160 A$\cdot$GeV. A description of the
transport approach is given in Refs. \cite{Cass99,Geiss} and the
production and propagation of charmed mesons is described in Refs.
\cite{new,Cass97,Geiss99}. The novel phenomenon addressed here
with respect to Ref. \cite{new} is the secondary production of
open charm mesons by 'meson'-'baryon' collisions with the new
parametrizations (\ref{sfit}) for the low-energy  $\pi N
\rightarrow \bar{D} (\bar{D}^*) \Lambda_c (\Sigma_c)$ processes.
We mention in passing that the role of secondary reactions on
intermediate mass dilepton pairs via the  $q \bar{q}\rightarrow
\gamma \rightarrow l^+l^-$  (Drell-Yan) mechanism has been
investigated in the UrQMD transport model in Ref. \cite{Spieles}
for nucleus-nucleus collisions at SPS energies before.

In Fig. 4 we show the open charm multiplicity (all $D$, $\bar{D},
D^*$, $\bar{D}^*, D_s^*$, $\bar{D}_s^*, D_s, \bar{D}_s$) as a
function of the impact parameter $b$ from the HSD approach for $Pb
+ Pb$ at 160 A$\cdot$GeV. Here the dashed line gives the yield
from primary baryon-baryon collisions, which is essentially the
yield from $pn$ reactions at $\sqrt{s} \approx$ 17 GeV times the
number of 'hard' $pN$ collisions (as given by Eq. (6) of Ref.
\cite{new}), while the solid line corresponds to the yield of open
charm from secondary reactions as described above.  Whereas for
peripheral collisions the secondary production is slightly smaller
than the direct channel, the yield from secondary meson-baryon
channels becomes larger with decreasing impact parameter $b$. The
tiny kinks in the curves in Fig. 4 are due to the limited
statistics of the transport calculations. Without explicit
representation we note that the final differential spectra of the
$D$-mesons in transverse momentum and rapidity are very similar to
those from primary and secondary channels, respectively, due to
rescattering with the surrounding hadrons (cf. Ref. \cite{new}).

In order to compare to the data from the NA50 Collaboration
\cite{NA50d} we have to adopt a model to convert the number of
participating nucleons $A_{part}$ to the impact parameter $b$ from
the transport calculation. We use
\begin{equation}
A_{part} =  2 A - N_0(b),
\end{equation}
where $N_0(b)$ stands for the number of noninteracting nucleons
(with no 'hard' collisions) from the transport code at impact
parameter $b$ while $A$ is the target (projectile) mass number.
Denoting the expected number of $D, \bar{D}$ mesons from primary
$BB$ collisions by $N_{prim.}$ and the number  from secondary
'meson'-'baryon' interactions by $N_{sec.}$ this leads to the
ratio
\begin{equation}
\label{ratio} R(b) =  \frac{N_{prim.} + N_{sec.}}{N_{prim.}}(b) =
R(A_{part}(b)) =  R(A_{part}) ,
\end{equation}
which is displayed in Fig. 5 (solid histogram) in comparison to
the data on the open charm enhancement from Ref. \cite{NA50d}. The
calculated ratio first increases fast with $A_{part}$ and becomes
almost constant for $A_{part} \geq 100$. This is due to the fact
that secondary semihard 'meson'-'baryon' reactions with 'wounded'
nucleons, that have scattered at least once, already set in for
large impact parameter $b$; their relative number then increases
only slightly faster than $\sim A_{part}$.
 Obviously, the general trend of the
data can be roughly described with increasing centrality for the
secondary cross sections specified above. The present statistical
uncertainty of the data \cite{NA50d} does not allow for a final
conclusion.

In summary: in this letter we have argued that for 'low' energy
'meson'-'baryon' reactions the dominant $c \bar{c}$ production is
related to the two-body (or quasi two-body) reactions $\pi N
\rightarrow \bar{D} (\bar{D^*})$  $\Lambda_c (\Sigma_c)$.
Estimates within the framework of the Quark-Gluon String model
suggest cross sections of a few $\mu b$ for $\pi N \rightarrow
\bar{D} \Lambda_c$ in the region of about 1 GeV above threshold.
The estimated order of magnitude for the open charm cross section
(cf. lower part of Fig. 3) is found to be compatible with the
'open charm enhancement' claimed by the NA50 Collaboration at the
SPS \cite{NA50d} without employing the assumption of thermal and
chemical or statistical equilibrium as advocated in Refs.
\cite{Gall,Rapp,Gorenstein}. It should be stressed, however, that
experimental investigations on open charm production in $\pi N$
reactions at invariant energies of 4.2 $\leq \sqrt{s} \leq $ 15
GeV are mandatory to confirm or disprove our suggestion.

We, furthermore, note that the cross section for charmonia such as
$J/\Psi$, $\chi_c$ or $\Psi^\prime$ are not substantially enhanced
by such secondary reaction channels since their cross sections are
small in $\pi N$ collisions \cite{Vogt99,new} such that no
substantial 'enhancement' of charmonia relative to the primary
$NN$ reaction channels is expected.

\vspace{1cm} \noindent The authors are grateful to E.L.
Bratkovskaya, A.B. Kaidalov and R. Vogt for clarifying
discussions.


\newpage

\begin{figure}[t]
\phantom{a}\vspace*{+2.5cm}
\centerline{\psfig{figure=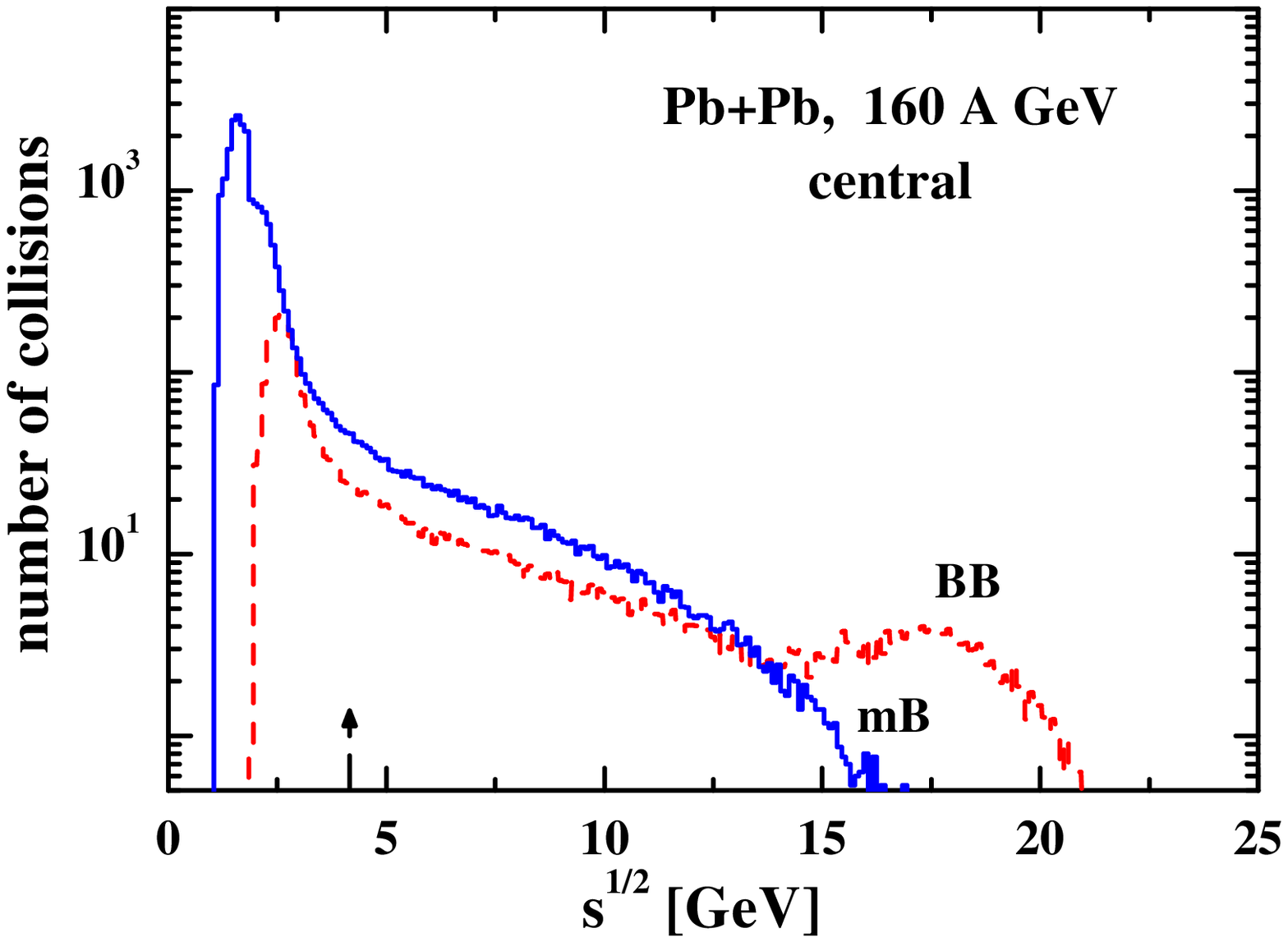,width=16cm}} \vspace*{-6cm}
\caption{The distribution in the invariant collision energy
$\sqrt{s}$ from primary baryon-baryon (dashed histogram; $BB$) and
secondary 'meson'-'baryon' interactions (solid histogram; $mB$)
for a central collision of $Pb + Pb$ at 160 A$\cdot$GeV from the
HSD transport approach. The arrow at $\sqrt{s} \approx$ 4.15 GeV
denotes the threshold for $\bar{D} + \Lambda_c$ production in
secondary interactions. Note, that contrary to Fig. 13 in Ref.
\cite{Cass00} 'meson' interactions with 'wounded' nucleons (or
diquarks) are taken into account, too. }
 \label{fig1}
\end{figure}

\begin{figure}[t]
\phantom{a}\vspace*{-2.5cm}
\centerline{\psfig{figure=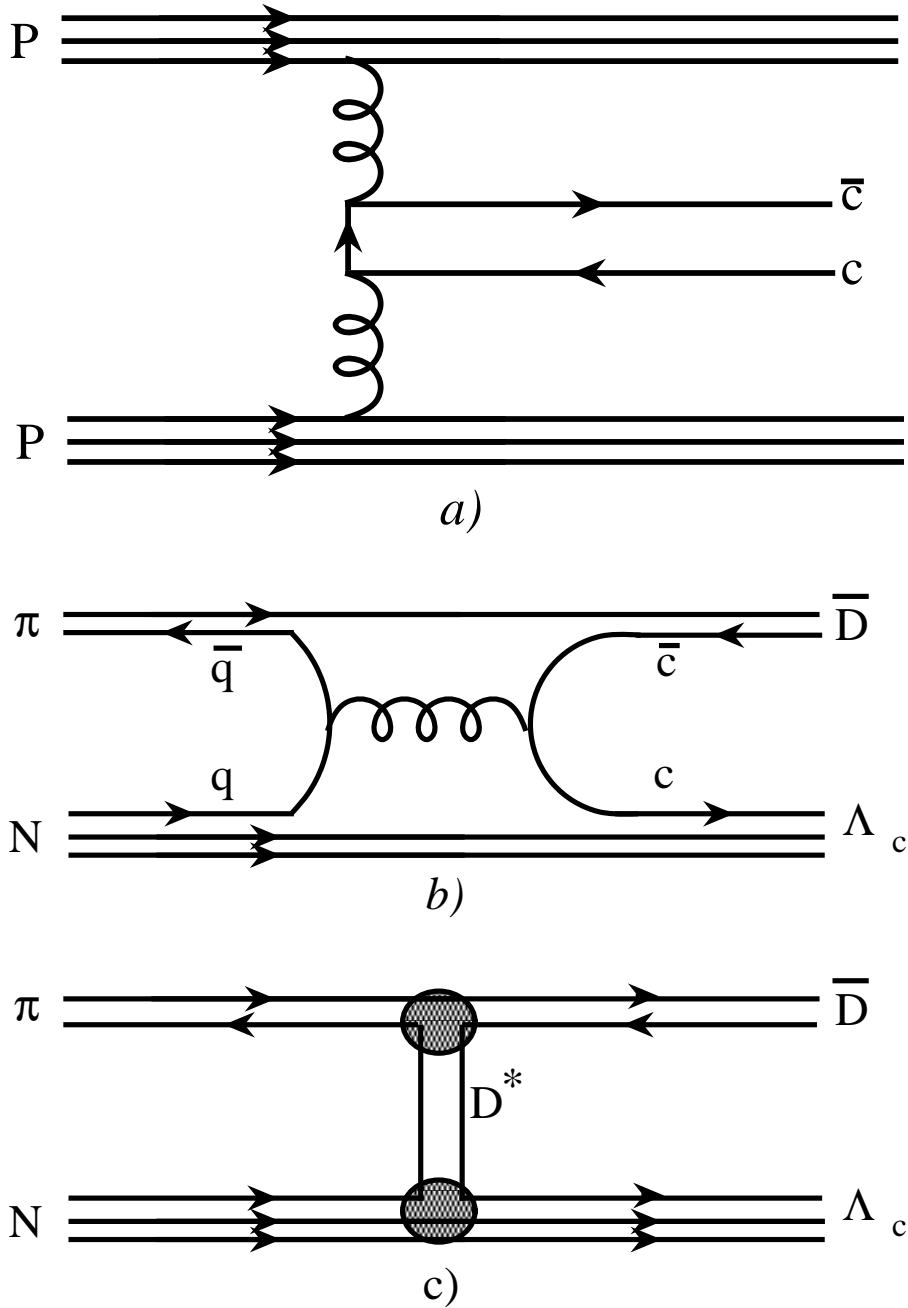,width=12cm}} \vspace*{+1cm}
\caption{Diagrams for $c \bar{c}$ production in $pp$ collisions
via two-gluon fusion (a), for $\bar{D} \Lambda_c$ production in
$\pi N$ collisions via $q \bar{q}$ annihilation (b) and for
$\bar{D} \Lambda_c$ production in $\pi N$ collisions via
$D^*$-exchange (c).}
 \label{fig2}
\end{figure}

\begin{figure}[t]
\phantom{a}\vspace*{-0.5cm}
\centerline{\psfig{figure=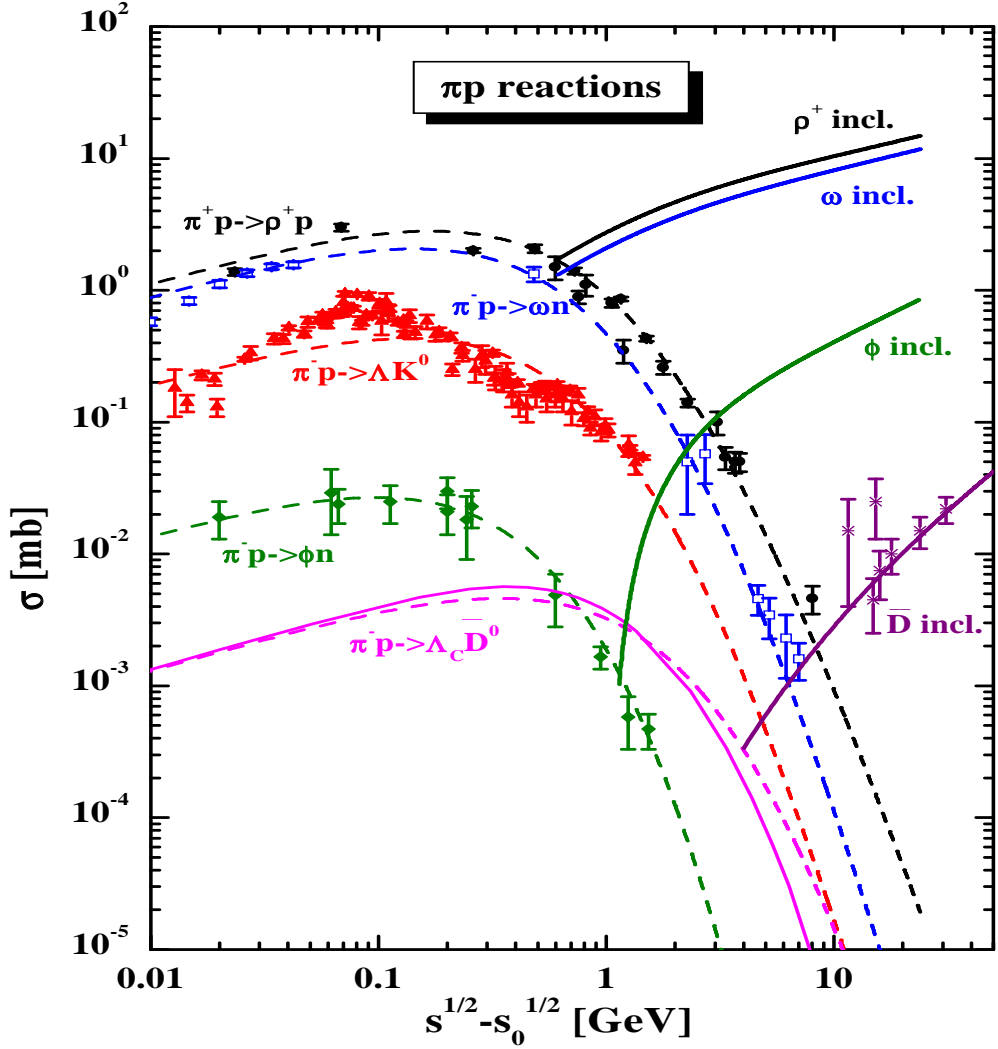,width=16cm,height=18cm}}
\vspace*{-3cm} \caption{The exclusive cross sections for the
reactions $\pi^+ p \rightarrow \rho^+ +p$, $\pi^- p \rightarrow
\omega + n$. $\pi^- p \rightarrow K^0 + \Lambda$ and $\pi^- + p
\rightarrow \phi + n$ from Ref. \cite{LB} in comparison to the
parametrizations (\ref{sfit}) (dashed lines). The inclusive cross
sections for $\rho^+, \omega$ and $\phi$ mesons are displayed in
terms of the upper thick solid lines, respectively, within the
parametrizations from the review \cite{Cass99}. The lower solid
line is the prediction for the process $\pi^- + p \rightarrow
\bar{D}^0 + \Lambda_c$ within the Regge model; the lower dashed
line is the corresponding parametrization by (\ref{sfit}).
 The inclusive yield
for all open charm mesons with a $\bar{c}$ quark ($\bar{D}$ incl.)
is shown in terms of the lower thick solid line  where the
functions (\ref{fit}) have been employed with the parameters from
Table 1. The related inclusive data ($*$) have been taken from
Ref. \cite{Tavernier}. }
 \label{fig3}
\end{figure}

\begin{figure}[t]
\phantom{a}\vspace*{0.5cm}
\centerline{\psfig{figure=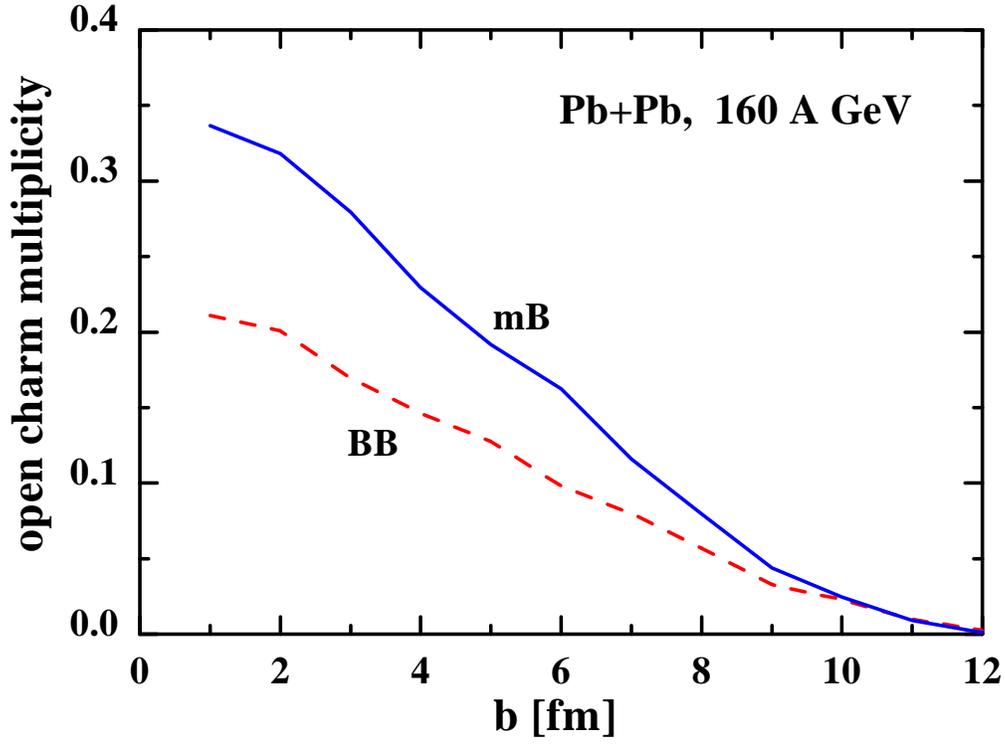,width=16cm}} \vspace*{-8cm}
\caption{The multiplicity of all $D, \bar{D}$ mesons from the HSD
transport approach for $Pb + Pb$ at 160 A$\cdot$GeV as a function
of impact parameter $b$. The dashed line stands for the yield from
primary baryon-baryon collisions while the solid line denotes the
contribution from secondary 'meson'-'baryon' collisions within the
parametrization (\ref{sfit}).}
 \label{fig4}
\end{figure}

\begin{figure}[t]
\phantom{a}\vspace*{0.5cm}
\centerline{\psfig{figure=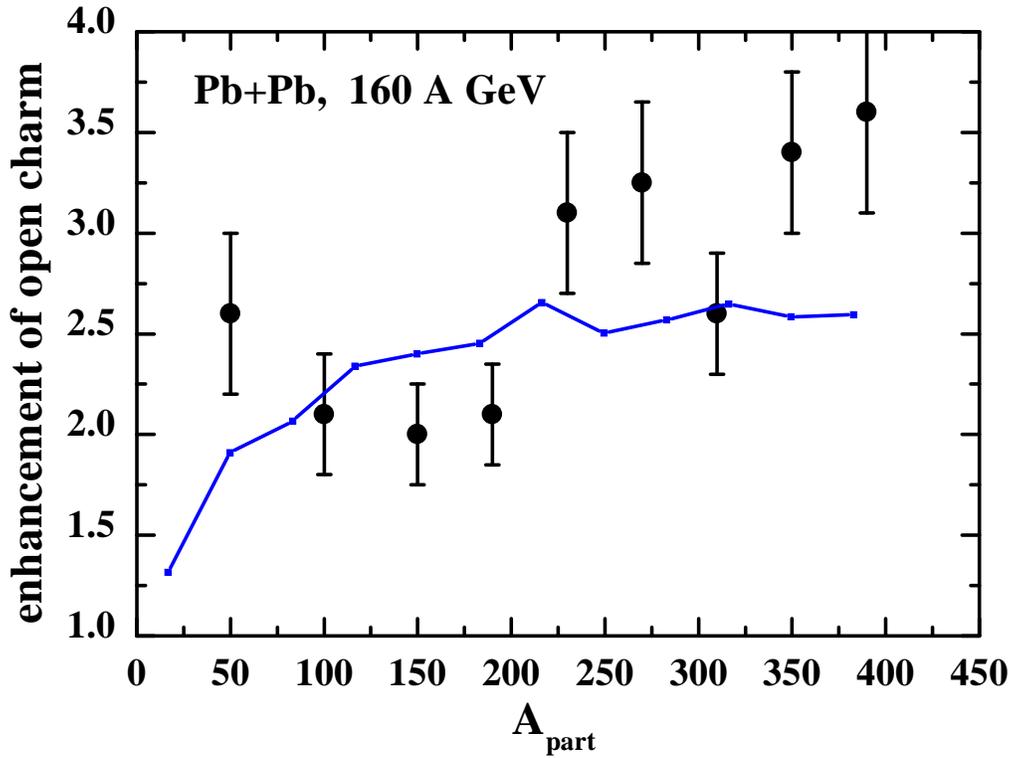,width=16cm}} \vspace*{-8cm}
\caption{The enhancement factor (\ref{ratio}) for the production
 of all $D, \bar{D}$ mesons from the HSD transport
approach for $Pb + Pb$ at 160 A$\cdot$GeV as a function of the
number of participants $A_{part}$ (see text) in comparison to the
data from Ref. \cite{NA50d}.}
 \label{fig5}
\end{figure}

\end{document}